\documentstyle[osa,manuscript]{revtex}
\begin{document}
\title{Evaluating the critical current magnitude and distribution width of
tridimensional Josephson junction arrays}
\author{W. A. C. Passos*, U. R. de Oliveira and W. A. Ortiz}
\address{Grupo de Supercondutividade e Magnetismo\\
Departamento de F\'{i}sica, Universidade Federal de S\~{a}o Carlos\\
Caixa Postal 676 - 13565-905 S\~{a}o Carlos, SP, Brazil}
\maketitle

\begin{abstract}
% DON'T CHANGE THIS LINE
In this contribution we present a simple and effective procedure to
determine the average critical current of a tridimensional disordered
Josephson junction array (3D-DJJA). Using a contactless configuration we
evaluate the average critical current and the typical width of the
distribution through the analysis of the isothermal susceptibility response
to the excitation field amplitude, $\chi _{AC}(h)$. A 3D-DJJA fabricated
from granular Nb is used to illustrate the method.
\end{abstract}

% INITIALIZE - DONT CHANGE
%
%
%

% \author{*pwac@df.ufscar.br}   % Use this and the next line only if there is a second
% \address{Another University, etc.}  % address. (Remove the left % marks)
%
\begin{verbatim}
*Corresponding Author: pwac@df.ufscar.br
\end{verbatim}

The low-frequency magnetic response of Nb-AlO$_{x}$-Nb Josephson junction
arrays (JJA) has been investigated by Araujo-Moreira and coworkers\cite
{araujo1}$^{,}$\cite{barbara2}, through the temperature $(T)$ and excitation
field $(h)$ dependence of the AC magnetic susceptibility, $\chi _{AC}$.
Based on simulations of a single-plaquette model, they have successfully
explained the dynamic reentrance presented by $\chi _{AC}(T)$ for certain
values of $h$, as well as the isothermal susceptibility $\chi _{AC}(h)$.
Simulated magnetization versus applied field curves for the single-plaquette
model, predict that a certain class of arrays would exhibit, upon excitation
by a field, a remanent moment in a limited interval of temperatures,
depending on the critical current, $I_{c}$, and other modeling parameters of
the junctions. As a matter of fact, the McCumber parameter, $\beta _{L}$ is
the actual key feature controlling the appearance of a magnetic remanence,
which is predicted to exist for $\beta _{L}<<1$\cite{araujo1}$^{,}$\cite
{barbara2}.

It has been demonstrated recently\cite{passos3} that tridimensional
disorderd Josephson junction arrays (3D-DJJA) can be produced in a
controlled manner. Arrays were fabricated from granular superconductors,
using either conventional (LTS) or high-temperature (HTS) powder. These
specimens have proved to exhibit\cite{passos3}$^{-}$\cite{passos5} all
relevant signatures of a JJA, including the typical Fraunhofer dependence of
the critical current with the applied magnetic field, the Wohlleben effect
(WE), and the magnetic remanence anticipated by numerical simulations.

On the other hand, these 3D-DJJAs can also be envisaged as especially
assembled specimens of granular superconductors and, conceivably, their
intragranular transport and magnetic properties could be treated by the
commonly employed approaches based on critical state models\cite{araujo6}$%
^{-}$\cite{chen10}. Evidently this would not be the case for the intergrain
response, since it is originated by weak links arranged in a 3D-DJJA, whose
behavior does not obey critical state models. In this contribution we
present results of a systematic study of the isothermal susceptibility
response to the excitation field amplitude, $\chi _{AC}(h)$, of a 3D-DJJA
fabricated from granular Nb.

Samples were fabricated following a standard procedure described elsewhere 
\cite{passos3}. In short, niobium powder is separated according to grain
size, using a set of special sieves, with mesh gauges ranging from 38 to 44 $%
\mu $m. The powder is then uniaxially pressed in a mold to form a
cylindrical pellet of 2.5 mm radius by 2 mm height. This pellet is a
tridimensional disordered JJA (3D-DJJA) in which the junctions are
weakly-coupled grains, i.e., weak-links formed by a sandwich between Nb
grains and a Nb-oxide layer originally present on the grain surface.

Measurements of $\chi _{AC}(h)$ were carried out using the AC-module of a
Quantum Design SQUID magnetometer, for an excitation frequency of 100 Hz, at
temperatures ranging from $T$ = 2 K up to $T_{c}$ = 8.9 K. In this paper we
focus on the upper part of the temperature window, closer to $T_{c}$. Fig.1
shows the real ($\chi $') and imaginary ($\chi $'') parts of $\chi _{AC}(h)$
for some values of $T$. The field at which $\chi $'' peaks, $h_{p}$, is an
indirect measure of the average critical current density of the
intergranular matrix\cite{araujo6}$^{-}$\cite{araujo8}, i.e., $<J_{c}>$ of
the array. For a sample of cylindrical shape of radius a, $h_{p}=a<J_{c}>$.
The exponential critical state model (ECSM)\cite{araujo7} was used to
simultaneously fit $\chi $' and $\chi $'', from which the temperature
dependence of $<J_{c}>$, its typical distribution width, $p(T)$, and the
granular fraction of the sample, $f_{g}(T)$, are determined. It is worth
mentioning that, as expected, the ECSM fits well the experimental data above
hp, but fails to fit the whole curve, as below hp the JJA behavior
substitutes that of a critical state. Consistently, up to $h=h_{p}$ the
array gives a positive (WE) contribution to the real part of $\chi _{AC}$,
so that $\chi $'$>-1$ and the sample is not perfectly diamagnetic. On the
other hand, the dispersive activity of the vortices differs from that of an
ordinary granular sample, being either lower, when the flux lines are
pinned, or higher, when they are temporarily free to relocate, depending on
the value of $h$.

The main curve in Fig.2 shows the average $J_{c}(T)$, whereas the insets
depict $p(T)$ (lower left) and $f_{g}(T)$ (upper right). The line connecting
the experimental points for the average $J_{c}(T)$ is a fit of the form $%
J_{c}(T)=J_{c0}(1-T/T_{c})^{2.38}$, as introduced by Wright and coworkers
for a matrix formed by grains linked by Josephson couplings\cite{wright11}$%
^{,}$\cite{wright12}. Here, $T_{c}$ is the critical temperature of the
array, which was obtained from the fitting as been 8.05 K, in excellent
agreement with the value of $T^{\ast }$ determined by independent means in
Ref. \cite{passos4} for the same sample. Not surprisingly, the numbers
obtained for the average critical current density of the 3D-DJJA are
comparable to those reported previously for the intergranular critical
current of a melt-textured YBa$_{2}$Cu$_{3}$O$_{7-\delta }$ sample\cite
{araujo7}, an ordered 2D-JJA of Nb-AlO$_{x}$-Nb\cite{araujo1} and a 3D-DJJA
of YBCO\cite{passos13}, among others. As could be anticipated, the critical
current distribution of the array broadens as $T$ approaches $T_{c}$, as can
be inferred by the continuous decrease of its typical dispersion, $p(T)$\cite
{araujo7}. A corresponding decrease on the granular fraction, measured by
the volume fraction of superconducting grains to the normal matrix, $f_{g}$,
occurs as the superconducting properties degrade with increasing $T$,
weakening at the grain boundaries and, from there, towards the center of the
grains.

To properly consider the significance of measuring $<J_{c}>$ of a 3D-DJJA
using a contactless configuration, one should bear in mind that performing
conventional current-voltage measurements in the disordered array studied
here, would be infeasible.

Financial support was partially provided by Brazilian agencies FAPESP,
CAPES, PRONEX and CNPq.

\begin{figure}[tbp]
\caption{Real ($\protect\chi $') and imaginary ($\protect\chi $'') parts of $%
\protect\chi _{AC}(h)$ for three values of $T$. The field at which $\protect%
\chi $'' peaks is an indirect measure of the average critical current
density of the array.}
\end{figure}

\begin{figure}[tbp]
\caption{Main curve: average critical current of the 3D-DJJA; lower left:
dispersion of critical current distribution, $p(T)$; upper right: granular
fraction, $f_{g}(T)$. Line connecting $J_{c}(T)$ points is a fit of the
expression $J_{c}(T)=J_{c0}(1-T/Tc)^{2.38}$. Lines on insets are only guides
to the eye.}
\end{figure}

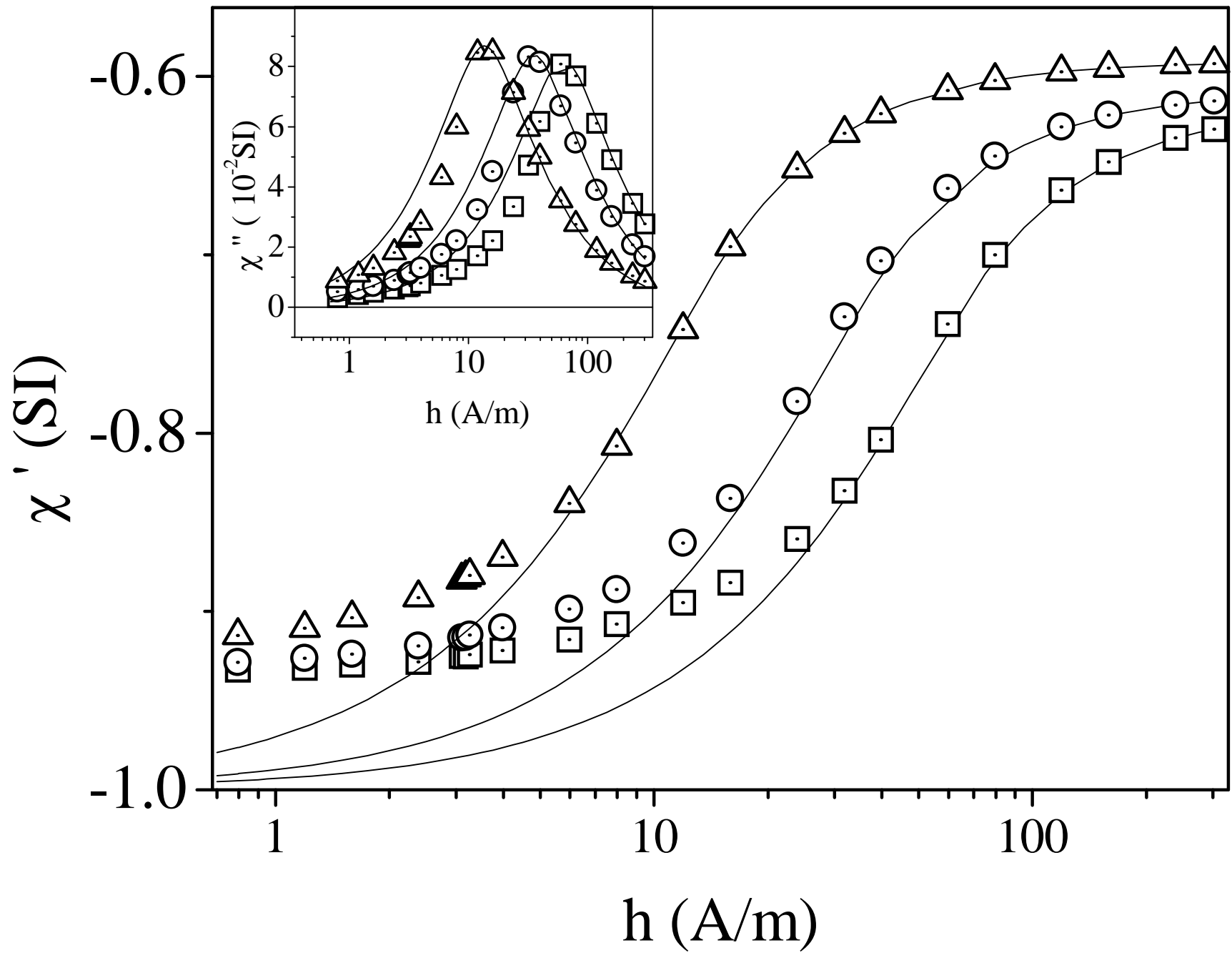
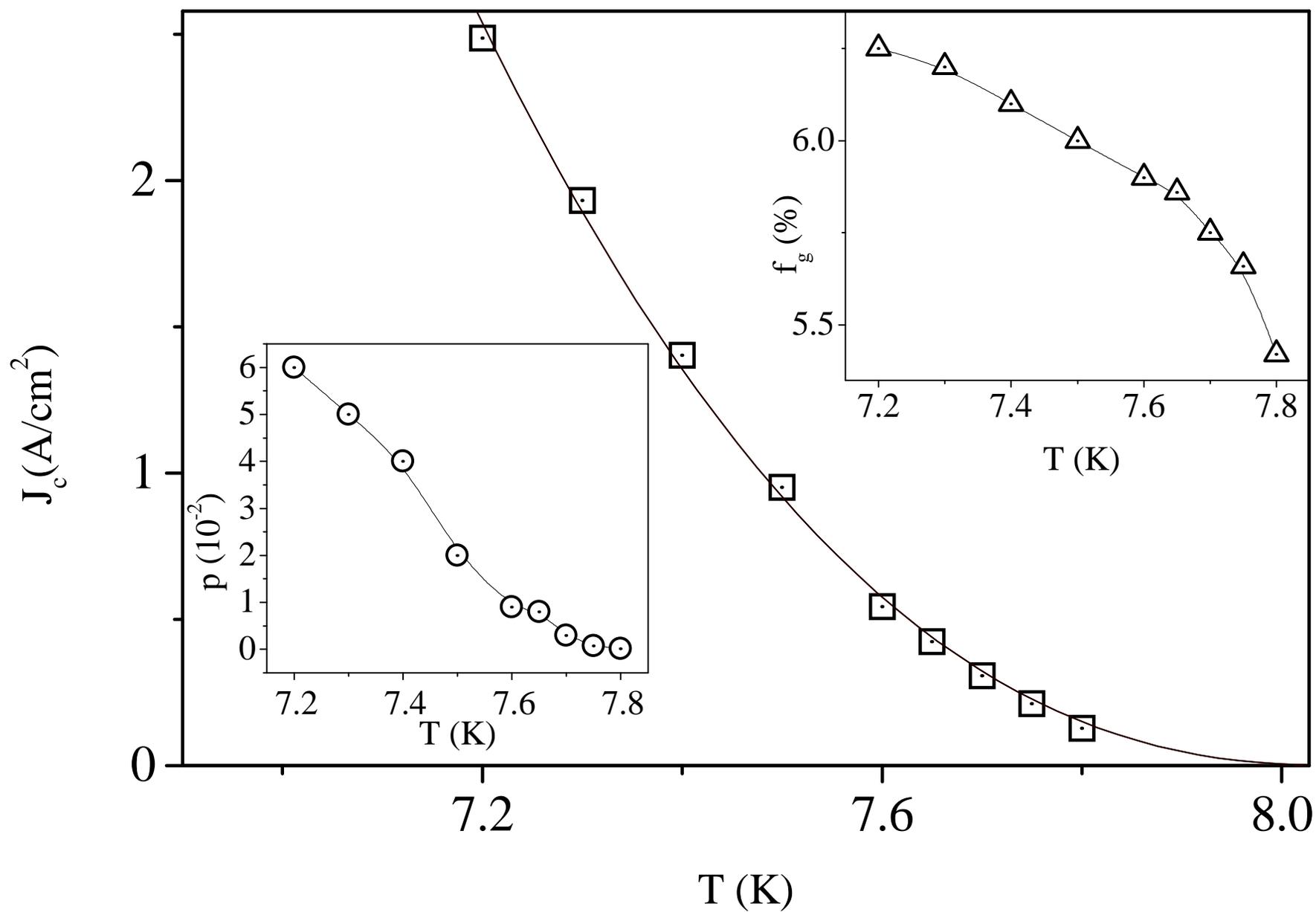
\end{document}